\documentclass[twocolumn,showpacs,preprintnumbers,amsmath,amssymb,showkeys]{revtex4}

\usepackage{graphicx}
\usepackage{dcolumn}
\usepackage{bm}
\usepackage{bezier}

\begin{document}

\title{Quantum Phase Slip Noise}
\author{Andrew G. Semenov$^{1,3}$
and Andrei D. Zaikin$^{2,1}$
}
\affiliation{$^1$I.E.Tamm Department of Theoretical Physics, P.N.Lebedev
Physical Institute, 119991 Moscow, Russia\\
$^2$Institute of Nanotechnology, Karlsruhe Institute of Technology (KIT), 76021 Karlsruhe, Germany\\
$^3$National Research University Higher School of Economics, 101000 Moscow, Russia
}

\begin{abstract}

Quantum phase slips (QPS) generate voltage fluctuations in superconducting nanowires. Employing Keldysh technique and making use of the phase-charge duality arguments we develop a theory of QPS-induced voltage noise in such nanowires. We demonstrate that quantum tunneling of the magnetic flux quanta across the wire yields quantum shot noise which obeys Poisson statistics and is characterized by a power law dependence of
its spectrum $S_\Omega$ on the external bias. In long wires $S_\Omega$ decreases with increasing frequency $\Omega$ and vanishes beyond a threshold value of $\Omega$ at $T \to 0$. Quantum coherent nature
of QPS noise yields non-monotonous dependence of $S_\Omega$ on $T$ at small $\Omega$.

\end{abstract}


\maketitle

Can a superconductor generate voltage fluctuations? More specifically, if an external bias is applied to a superconductor could the latter produce shot noise?
Posing these questions we, of course, imply that temperature $T$, characteristic frequencies and/or voltages as well as all other
relevant energy parameters remain well below the superconducting gap, i.e. the superconductor is either in or
sufficiently close to its quantum ground state.

At the first sight, positive answers to both these questions can be rejected on fundamental grounds.
Indeed, a superconducting state is characterized by zero resistance, i.e. a non-dissipative current below some critical value can pass through the system.
Hence, neither non-zero average voltage nor voltage fluctuations can be expected.

These simple considerations -- although applicable to bulk superconductors -- become insufficient in the case of ultrathin superconducting wires
because of the presence of quantum phase slips (QPS) \cite{AGZ,Bezr08,Z10,Bezrbook}. In such wires quantum fluctuations of the superconducting order
parameter field $\Delta=|\Delta|e^{i\varphi}$ play an important role being responsible for temporal local suppression
of $|\Delta|$ inside the wire and, hence, for the phase slippage process. Each quantum phase slip event corresponds to the net phase jump by
$\delta \varphi =\pm 2\pi$ implying positive or negative voltage pulse  $\delta V=\dot{\varphi}/2e$ (here and below we set $\hbar =1$)
and tunneling of one magnetic flux quantum $\Phi_0\equiv \pi/e =\int |\delta V(t)|dt$ across the wire in the direction perpendicular
to its axis. Biasing the wire by an external current $I$ one breaks the symmetry between positive and negative voltage pulses
making the former more likely than the latter. As a result, the net voltage drop $V$ occurs across the wire also implying non-zero
resistance $R=V/I$ which may not vanish down to lowest $T$ \cite{ZGOZ,GZQPS}, as it was indeed observed in a number of experiments  \cite{BT,Lau,Zgi08}.
Hence, in the presence of QPS the current flow becomes dissipative
and -- according, e.g., to the fluctuation-dissipation theorem (FDT) -- one should also expect voltage fluctuations to occur in the system.

While these arguments suggest a positive answer to the first of the above questions they do not yet specifically address
shot noise. Two key pre-requisits of shot noise are:  ($i$) the presence of discrete
charge carriers (e.g., electrons) in the system and ($ii$) scattering of such carriers at disorder.
 Although discrete charge carriers -- Cooper pairs --
are certainly present in superconducting nanowires, they form a superconducting condensate
flowing along the wire {\it without any scattering}. For this reason the possibility for shot noise to occur in superconducting
nanowires appears by no means obvious.

In this Letter we will perform a detailed theoretical analysis of QPS-induced voltage fluctuations in ultrathin superconducting wires. In particular,
we will demonstrate that
quantum tunneling of magnetic flux quanta $\Phi_0$ across the wire causes shot noise which obeys Poisson statistics and shows a
non-trivial dependence on temperature, frequency and external current.

{\it The model and effective Hamiltonian}. The system under consideration is displayed in Fig. 1.
It consists of an ultrathin superconducting wire of length $L$ and cross section $s$ and a capacitance $C$
switched in parallel to this wire. The right end of the wire ($x=L$) is grounded as shown in the figure
($x$ is the coordinate along the wire ranging from 0 to $L$). The voltage $V(t)$ at its left end $x=0$
fluctuates  and such fluctuations can be measured by a detector. The whole system is biased by an external current $I=V_x/R_x$.

\begin{figure}[h]
\includegraphics[width=0.89\columnwidth]{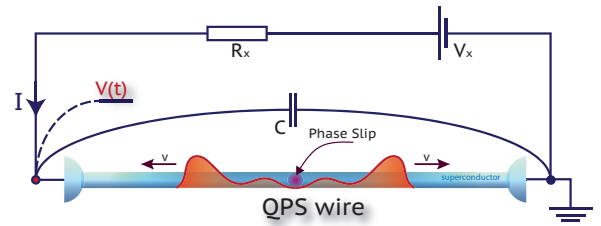}
\caption{(Color online) The system under consideration. The figure also illustrates creation of two plasmons by a QPS.}
\label{fig1}
\end{figure}

An effective Hamiltonian for our system can be written in the form
\begin{equation}
\hat H = \hat H_{Ch}-I\varphi /2e + \hat H_{\rm wire}.
\label{Ham}
\end{equation}
The first and the second terms in the right-hand side of Eq. (\ref{Ham}) account respectively for the charging energy \cite{SZ90}
\begin{equation}
\hat H_{Ch}= \frac{1}{2C}\left(-i \frac{\partial}{\partial (\varphi/2e)} +Q\right)^2
\label{Hch}
\end{equation}
and for the potential energy tilt produced by an external current $I$. The variable $\varphi (t) \equiv \varphi (0,t)$ represents the phase of
the superconducting order parameter field $\Delta (x,t)$ at $x=0$. Here we also set $\varphi (L,t)\equiv 0$.

The last term $\hat H_{\rm wire}$ in Eq. (\ref{Ham}) describes the superconducting wire. This part of the effective Hamiltonian can be
expressed in terms of both the modulus $|\Delta (x,t)|$ and the phase $\varphi (x,t)$ of the order parameter field \cite{ZGOZ,GZQPS,ogzb}. Here, however,
we will proceed differently and employ the duality arguments.

The duality between the phase and the charge variables was established and discussed in details in the case
of ultrasmall Josephson junctions \cite{SZ90,PZ88,averin,Z90}. Later the same duality arguments were extended to short \cite{MN} and
long \cite{S10,HN,SZ13} superconducting wires. According to the results
\cite{SZ13} the dual representation for the Hamiltonian of a superconducting nanowire is defined by an effective sine-Gordon model
\begin{equation}
\hat H_{\rm wire} = \hat H_{TL}+ \hat H_{QPS}.
\label{Hamw}
\end{equation}
In the absence of quantum phase slips such nanowire
can be described as a transmission line with
\begin{equation}
\hat H_{TL}=\int_0^Ldx \left(\frac{\Phi^2}{2\mathcal{L}_{\rm kin}}+\frac{(\partial_x\chi )^2}{2C_{\rm w}\Phi_0^2}\right),
\end{equation}
where $\mathcal{L}_{\rm kin}=1/(\pi\sigma_N\Delta_0 s)$ and $C_{\rm w}$ are respectively the kinetic wire inductance (times length) and the geometric wire capacitance (per length),
\begin{equation}
[\Phi (x),\chi (x')]=-i\Phi_0\delta (x-x')
\end{equation}
defines the commutation relation between the canonically conjugate flux (or phase) and charge operators, $\sigma_N$ is the normal state Drude
conductance of the wire and $\Delta_0$ is the superconducting gap. The term
\begin{equation}
\hat H_{QPS}=-\gamma_{QPS}\int_0^Ldx \cos \chi
\label{HQPS}
\end{equation}
accounts for the effect of quantum phase slips and
\begin{equation}
\gamma_{QPS} \sim (g_\xi\Delta_0/\xi)\exp (-ag_\xi), \quad a \sim 1,
\label{gQPS}
\end{equation}
is the QPS tunneling amplitude \cite{GZQPS} per unit wire length with $g_\xi =2\pi\sigma_N s/(e^2\xi) \gg 1$ being the dimensionless normal state conductance of the
wire segment of length equal to the coherence length $\xi$.

The physical meaning of the quantum field $\chi (x,t)$ is transparent: It is proportional to the total charge $q(x,t)$ that has passed through
the point $x$ up to the time moment $t$, i.e. $q(x,t)=\chi (x,t)/\Phi_0$. Accordingly, the local current $I(x,t)$ and the
local charge density $\rho (x,t)$ are defined as
\begin{equation}
I(x,t)=\partial_t\chi (x,t)/\Phi_0, \quad \rho (x,t)=-\partial_x\chi (x,t)/\Phi_0,
\end{equation}
thereby satisfying the continuity equation. The charge $Q$ in Eq.
(\ref{Hch}) equals to $Q(t)=\chi (0,t)/\Phi_0$.

{\it Keldysh technique and perturbation theory}. In order to proceed we will make use of the Keldysh path integral technique. Accordingly,
our variables of interest
need to be defined on the forward and backward time branches of the Keldysh contour, i.e. we now have $\varphi_{F,B} (t)$ and $\chi_{F,B}(x,t)$.
As usually, it is convenient to also introduce the ``classical'' and ``quantum'' variables, respectively
$\varphi_+ (t)= (\varphi_F (t)+\varphi_B (t))/2$ and $\varphi_- (t)= \varphi_F (t)-\varphi_B (t)$ (and similarly for the $\chi$-fields).
Making use of the Josephson relation between the voltage and the phase one can formally express the expectation value of the voltage operator
across the the superconducting wire in the form
\begin{equation}
\langle V(t_1)\rangle=\frac{1}{2e}\left\langle\dot \varphi_{+}(t_1) e^{iS_{QPS}}\right\rangle_0
\label{V}
\end{equation}
where
\begin{equation}
 S_{QPS}=-2 \gamma_{QPS}\int dt\int\limits_0^L dx\sin(\chi_{+})\sin (\chi_-/2)
\end{equation}
and
\begin{equation}
\langle ...\rangle_0 =\int\mathcal D^2\varphi (t) \mathcal D^2\chi (x,t) (...) e^{iS_0[\phi , \chi ]}
\end{equation}
implies averaging with the Keldysh effective action $S_0$ corresponding to the Hamiltonian $\hat H_{0} = \hat H- \hat H_{QPS}$. Analogously, for
the voltage-voltage correlator $\langle V(t_1)V(t_2)\rangle =\frac12\langle\{\hat V(t_1),\hat V(t_2)\}  \rangle$ (where curly brackets denote the anticommutator) one has
\begin{equation}
\langle V(t_1)V(t_2)\rangle =\frac{1}{4e^2}\left\langle\dot \varphi_{+}(t_1)\dot \varphi_{+}(t_2) e^{iS_{QPS}}\right\rangle_0,
\label{VV}
\end{equation}
Higher voltage correlators are defined similarly. Their analysis, however, is beyond the frames of this work.

Eqs. (\ref{V}) and (\ref{VV}) are formally exact expressions which we are now going to evaluate. To this end we will employ the regular perturbation theory in $\gamma_{QPS}$ (\ref{gQPS}) which can be regarded as a small parameter of our theory. In the zero order in $\gamma_{QPS}$ the problem is described by the quadratic (in both $\varphi$ and $\chi$) Hamiltonian $\hat H_0$ and all averages can be handled exactly with the aid of the Green functions
\begin{eqnarray}
G^R_{ab}(X,X')=-i\langle a_{+}(X)b_-(X')\rangle , \nonumber\\
G^K_{ab}(X,X')=-i\langle a_{+}(X)b_{+}(X')\rangle ,
\label{GRK}
\end{eqnarray}
where $a(X)$ and $b(X)$ stand for one of the fields $\varphi (t)$ and $\chi (x,t)$. As both these fields are
real, the advanced  and retarded Green functions obey the condition $G^A_{ab}(\omega)=G^R_{ba}(-\omega)$.
With this in mind the Keldysh function $G^K$ can be expressed in the form
\begin{eqnarray}
 G^K_{ab}(\omega)=\frac12\coth\left(\frac{\omega}{2T}\right)\left(G^R_{ab}(\omega)-G^R_{ba}(-\omega)\right).
\label{GK}\end{eqnarray}

Expanding Eqs. (\ref{V}) and (\ref{VV}) up to the second order in $\gamma_{QPS}$ and performing all necessary averages
we evaluate the results in terms of the Green functions (\ref{GRK}), see Supplemental material for further details.
The result of our calculation both for the average voltage (\ref{V}) and for the voltage-voltage correlator (\ref{VV}) can
also be expressed in the form of ``candy'' diagrams displayed in Fig. 2.

\begin{figure}[h]
\includegraphics[width=0.89\columnwidth]{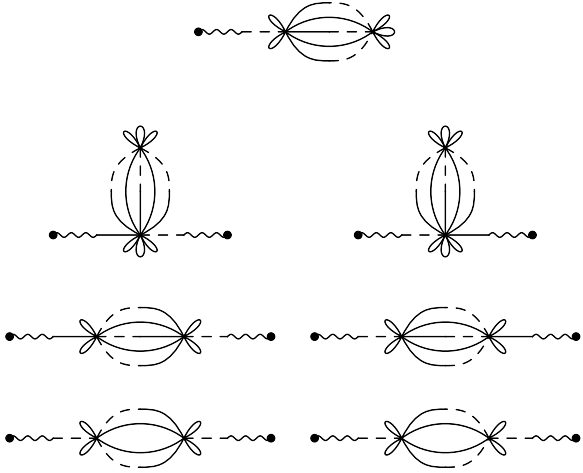}
\caption{Candy-like diagrams which determine both average voltage (\ref{V}) (upper diagram) and voltage noise (\ref{VV}) (six remaining diagrams) in the second order in $\gamma_{QPS}$. The fields $\varphi_+$, $\chi_+$ and $\chi_-$ in the propagators (\ref{GRK}) are denoted respectively by wavy, solid and dashed lines.}
\label{fig2}
\end{figure}

{\it I-V curve and voltage noise}. To begin with, let us briefly re-derive the results \cite{ZGOZ}
for the average voltage within the framework of our technique. We obtain
\begin{multline}
 \langle V\rangle=\frac{i\gamma_{QPS}^2}{4e}\int\limits_{0}^L dx\int\limits_0^L dx'\left(\lim_{\omega\to 0}\omega G_{\varphi\chi}^R(x;\omega)\right)
 \\
 \times\left({\mathcal P}_{x,x'}(-\Phi_0 I)-{\mathcal P}_{x,x'}(\Phi_0I)\right),
\label{V1}
\end{multline}
where ${\mathcal P}_{x,x'}(\omega)=P_{x,x'}(\omega)+\bar P_{x,x'}(\omega)$ and
\begin{eqnarray}
 \label{P}
 P_{x,x'}(\omega)=\int\limits_{0}^\infty dt e^{i\omega t}e^{i{\mathcal G}(x,x';t,0)},\\
{\mathcal G}(x,x';t,0)= G^K_{\chi\chi}(x,x';t,0)-\frac 12G^K_{\chi\chi}(x,x;t,t)\nonumber\\
-\frac 12G^K_{\chi\chi}(x',x';0,0)+\frac 12 G^R_{\chi\chi}(x,x';t,0).\nonumber
\end{eqnarray}
Bearing in mind that $\lim_{\omega\to 0}\omega G_{\varphi\chi}^R(x;\omega)=2\pi i$, Eq. (\ref{V1})
can be cast to the form
\begin{equation}
\langle V \rangle =\Phi_0\left(\Gamma_{QPS}(I)-\Gamma_{QPS}(-I)\right),
\label{V2}
\end{equation}
where we identify $\Gamma_{QPS}$ as
\begin{equation}
 \Gamma_{QPS}(I)=\frac{\gamma_{QPS}^2}{2}\int\limits_{0}^L dx\int\limits_0^L dx'{\mathcal P}_{x,x'}(\Phi_0I).
\end{equation}
Comparing the result (\ref{V2}) with that found in Ref. \onlinecite{ZGOZ} we immediately conclude that
$\Gamma_{QPS}(I)$ defines the quantum decay rate of the current state due to QPS. In \cite{ZGOZ} this rate was evaluated from the imaginary
part of the free energy $\Gamma_{QPS}(I)=2{\rm Im}F$. Here we derived the expression for $\Gamma_{QPS}$ by means of the real time technique
without employing the ${\rm Im}F$-method.

Making use of the above results, evaluating the Green functions (\ref{GRK}) (see Supplemental material) and keeping in mind the detailed balance condition
\begin{equation}
{\mathcal P}_{x,x'}(\omega)=e^{\frac{\omega}{T}}{\mathcal P}_{x,x'}(-\omega)
\label{dbc}
\end{equation}
we obtain
\begin{equation}
\langle V\rangle =\frac{\Phi_0 L v \gamma_{QPS}^2}{2}\varsigma^2\left(\frac{\Phi_0 I}{2}\right)\sinh\left(\frac{\Phi_0 I}{2T}\right),
\label{V3}
\end{equation}
where $v=1/\sqrt{{\mathcal L}_{\rm kin}C_{\rm w}}$ is the plasmon velocity \cite{ms},
\begin{equation}
\varsigma(\omega)=\tau_0^\lambda (2\pi T)^{\lambda -1}\frac{\Gamma\left(\frac{\lambda}{2}-\frac{i\omega}{2\pi T}\right)\Gamma\left(\frac{\lambda}{2}+\frac{i\omega}{2\pi T}\right)}{\Gamma(\lambda)},
\label{vsi}
\end{equation}
$\tau_0 \sim 1/\Delta_0$ is the QPS core size in time and $\Gamma (x)$ is the
Gamma-function. Here we also introduced the parameter \cite{ZGOZ} $\lambda=R_q/2Z_{\rm w}\propto \sqrt{s}$, where $R_q =\pi/2e^2$ is the "superconducting" quantum resistance unit and $Z_{\rm w}=\sqrt{\mathcal{L}_{\rm kin}/C_{\rm w}}$ is the wire impedance. It is satisfactory to observe that the result (\ref{V3}), (\ref{vsi}) matches with that found in Ref. \onlinecite{ZGOZ} by means of a different technique \cite{FN0}.

Let us now turn to voltage fluctuations. Our perturbative analysis allows to recover three different contributions
to the noise power spectrum, i.e.
\begin{equation}
 S_{\Omega}=\int dt e^{i\Omega t}\langle V(t)V(0)\rangle =S^{(0)}_{\Omega}+S^{r}_{\Omega}+S^{a}_{\Omega}.
\label{VV2}
\end{equation}
The first of these contributions $S^{(0)}_{\Omega}$ has nothing to do with QPS and just defines equilibrium voltage noise for a transmission line.
It reads
\begin{equation}
S^{(0)}_{\Omega}= \frac{i\Omega^2\coth\left(\frac{\Omega}{2T}\right)}{16e^2}\left(G_{\varphi\varphi}^R(\Omega)-
G_{\varphi\varphi}^R(-\Omega)\right).
 \end{equation}
The other two terms are due to QPS effects. The term $S^{r}_{\Omega}$ is also proportional to $\coth\left(\frac{\Omega}{2T}\right)$ and contains
the products of two retarded (advanced) Green functions:
\begin{multline}
S^{r}_{\Omega}=\frac{\gamma_{QPS}^2\Omega^2\coth\left(\frac{\Omega}{2T}\right)}{8e^2}\int\limits_{0}^L dx\int\limits_0^L dx' {\rm Re}\left[G_{\varphi\chi}^R(x;\Omega) \right.\\
 \left.\times ({\mathcal F}_{x,x'}(\Omega )G_{\varphi\chi}^R(x';\Omega)
 -{\mathcal F}_{x,x'}(0)G_{\varphi\chi}^R(x;\Omega))\right].
 \label{Sr}
 \end{multline}
Here we denoted
\begin{multline}
{\mathcal F}_{x,x'}(\Omega )=
- P_{x,x'}(\Omega +\Phi_0I)-P_{x,x'}(\Omega -\Phi_0I)\\
+\bar P_{x,x'}(-\Omega +\Phi_0I)+\bar P_{x,x'}(- \Omega -\Phi_0I).
\end{multline}
The remaining term $S^{a}_{\Omega}$, in contrast, contains the product of one retarded and one advanced Green functions and scales with
the combinations ${\mathcal C}_{\pm} = \coth\left(\frac{\Omega\pm\Phi_0 I}{2T}\right)-\coth\left(\frac{\Omega}{2T}\right)$ as
 \begin{eqnarray}
\label{Sa}
S^{a}_{\Omega}=\frac{\gamma_{QPS}^2\Omega^2}{16e^2}\int\limits_{0}^L dx\int\limits_0^L dx'G_{\varphi\chi}^R(x;\Omega)G_{\varphi\chi}^R(x';-\Omega)\\
\times \left[\sum_{\pm}{\mathcal C}_{\pm}\left(
 {\mathcal P}_{x,x'}(\Omega\pm\Phi_0I)-{\mathcal P}_{x,x'}(-\Omega\mp\Phi_0I)\right)\right].\nonumber
\end{eqnarray}
Eqs. (\ref{VV2})-(\ref{Sa}) together with the expressions for the Green functions (see Supplemental material) fully determine the voltage
noise power spectrum of a superconducting nanowire in the perturbative in QPS regime and represent the central result of this work.

In the zero bias limit $I\to 0$
the term $S^{a}_{\Omega}$ vanishes, and the equilibrium noise spectrum $S_{\Omega}=S^{(0)}_{\Omega}+S^{r}_{\Omega}$ is determined from FDT, see also \cite{SZ13}.
At non-zero bias values the QPS noise turns non-equilibrium. In the zero frequency limit $\Omega \to 0$
the terms $S^{(0)}_{\Omega}$ and $S^{r}_{\Omega}$ tend to zero, and the voltage noise $S_{\Omega \to 0}\equiv S_0$ is determined solely by $S^{a}_{\Omega}$. Then from Eq. (\ref{Sa}) we obtain
\begin{multline}
 S_{0}= \Phi_0^2\left(\Gamma_{QPS}(I)+\Gamma_{QPS}(-I)\right)\\
 =\Phi_0\coth\left(\frac{\Phi_0 I}{2T}\right)\langle V \rangle,
\label{shot1}
\end{multline}
where $\langle V \rangle$ is specified in Eqs. (\ref{V2}), (\ref{V3}).  Combining
the result (\ref{shot1}) with Eqs. (\ref{V3}), (\ref{vsi}) we find
\begin{equation}
\label{S0lim}
S_{0}\propto
\begin{cases}
T^{2\lambda -2}, & T\gg \Phi_0I,
\\
I^{2\lambda -2}, & T\ll \Phi_0I.
\end{cases}
\end{equation}
At higher temperatures $T\gg \Phi_0I$ (though still $T \ll \Delta_0$) Eq. (\ref{S0lim})
just describes equilibrium voltage noise $S_0=2TR$ of a linear Ohmic resistor $R=\langle V \rangle /I \propto T^{2\lambda -3}$ \cite{ZGOZ}.
In the opposite low temperature limit $T\ll \Phi_0I$ it accounts for QPS-induced {\it shot noise} $S_0=\Phi_0\langle V \rangle$ obeying
{\it Poisson statistics} with an effective ``charge'' equal to the flux quantum $\Phi_0$.

This result sheds light on
the physical origin of shot noise in superconducting nanowires: It is produced by coherent tunneling of magnetic flux quanta $\Phi_0$ across the wire.
In the dual picture \cite{SZ13} such flux quanta can be viewed as charged quantum particles passing through (and being scattered at) an effective spatially extended tunnel barrier.

Note that previously the result analogous to Eq. (\ref{shot1}) was derived for thermally activated phase slips (TAPS) \cite{GZTAPS}.
This similarity appears remarkable given a crucial physical difference between TAPS and QPS: The former can be regarded as classical (i.e. incoherent) and non-interacting objects, whereas the latter are fully coherent \cite{Ast} forming an interacting quantum gas.

Another interesting limiting case is that of sufficiently high frequencies  and/or long wires $v/L\ll\Omega\ll \Delta_0$. In this limit we obtain
\begin{equation}
S^{(0)}_{\Omega}=\frac{\lambda}{8\pi e^2}\frac{\Omega\coth\left(\frac{\Omega}{2T}\right)}{(\Omega/2E_C)^2+(\lambda /\pi )^2}.
\label{S0L}
\end{equation}
This contribution is independent of the wire length $L$. At low $T$ and $\Omega /\lambda \gtrsim  E_C=e^2/2C$ we have $S^{(0)}_{\Omega} \propto 1/\Omega$, i.e. the wire may generate $1/f$ voltage noise.
Evaluating the QPS terms $S^{r}_{\Omega}$ and $S^{a}_{\Omega}$ we observe that the latter scales linearly with the wire length $L$ whereas the former does not. Hence,
the term $S^{r}_{\Omega}$ can be safely neglected in the long wire limit. For the remaining QPS term $S^{a}_{\Omega}$ we get
\begin{multline}
S^{a}_{\Omega}=\frac{L\lambda^2v\gamma_{QPS}^2}{4e^2}\left[\varsigma\left(\frac{\Phi_0 I}{2}-\Omega\right)
-\varsigma\left(\frac{\Phi_0 I}{2}+\Omega\right)\right]\\
\times\frac{\sinh\left(\frac{\Phi_0 I}{2T}\right)\varsigma\left(\frac{\Phi_0 I}{2}\right)}{\left((\Omega/2E_C)^2+(\lambda /\pi)^2\right)\sinh\left(\frac{\Omega}{2T}\right)}.
\label{Sa3}
\end{multline}
At $T\to 0$ from Eq. (\ref{Sa3}) we find
\begin{equation}
S^{a}_{\Omega}\propto
\begin{cases}
I^{\lambda -1}(I-2\Omega /\Phi_0)^{\lambda -1}, & \Omega < \Phi_0I/2,
\\
0, & \Omega > \Phi_0I/2.
\end{cases}
\label{Sa4}
\end{equation}
This result can be interpreted as follows.
At $T=0$ each QPS event excites (at least) two plasmons \cite{FN} (see Fig. 1) with total energy $E=\Phi_0 I$ propagating in the
opposite directions along the wire. One plasmon (with energy $E/2$) gets dissipated at the grounded end of the wire
while another one (also with energy $E/2$) reaches its opposite end causing voltage fluctuations (emits a photon) with frequency $\Omega$ measured by a detector. Clearly, at $T=0$ this process is only possible at
$\Omega < E/2$ in the agreement with Eq. (\ref{Sa4}).
\begin{figure}[h]
\includegraphics[width=0.89\columnwidth]{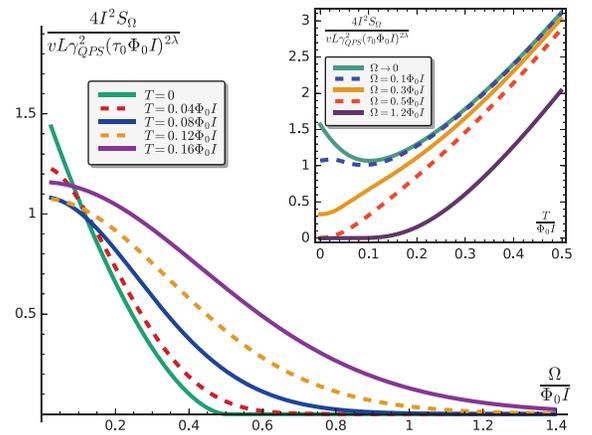}
\caption{(Color online) The frequency dependence of the QPS noise spectrum $S_{\Omega}$ (\ref{Sa3}) at $ \lambda =2.7$, large $E_C$ and different $T$ in the long wire limit. The inset shows $S_{\Omega}$ as a function of $T$.}
\label{fig3}
\end{figure}

The result (\ref{Sa3}) is also illustrated in Fig. 3. At sufficiently small $\Omega$ (we still keep $\Omega \gg v/L$) one observes a non-monotonous dependence of $S_{\Omega}$ on $T$ which is a direct consequence
of quantum coherent nature of QPS noise.

Finally, we point out that the perturbative in $\gamma_{QPS}$ approach employed here is fully justified for
not too thin wires with $\lambda > \lambda_c \simeq 2$ \cite{ZGOZ}. In wires with  $\lambda < \lambda_c$
(characterized by unbound QPS-anti-QPS pairs) $\gamma_{QPS}$ gets effectively renormalized to higher values and, hence,
the perturbation theory eventually becomes obsolete. However, even in this case our results may still remain applicable at sufficiently high temperature, frequency and/or current values. In the low energy limit
long wires with $\lambda <\lambda_c$ show an insulating behavior, as follows from the exact solution of the
corresponding sine-Gordon model \cite{CET}. This solution suggests that also voltage fluctuations become large in this limit.

In summary, we demonstrated that quantum phase slips generate voltage noise in superconducting nanowires. In the presence of a current bias $I$
quantum tunneling of the magnetic flux quanta $\Phi_0$ across the wire causes Poissonian shot noise with a non-trivial power law dependence of
its spectrum on both $I$ and frequency $\Omega$. Our predictions can be directly verified in future experiments and need to be observed
while optimizing the operation of QPS qubits \cite{MH}.

We acknowledge useful discussions with K.Yu. Arutyunov, D.S. Golubev and P. Hakonen.

\begin{widetext}

\clearpage

\newpage
\begin{center}
{\LARGE \textbf{Supplemental Material}}
\end{center}
\setcounter{equation}{0}
\setcounter{figure}{0}
\setcounter{table}{0}
\setcounter{page}{1}
\makeatletter
\renewcommand{\theequation}{S\arabic{equation}}
\renewcommand{\thefigure}{S\arabic{figure}}
\renewcommand{\bibnumfmt}[1]{[S#1]}
\renewcommand{\citenumfont}[1]{S#1}
\subsection{Perturbation theory}

Let us expand the general expressions (\ref{V}) and (\ref{VV}) up to the second order in $\gamma_{QPS}$.
It is easy to demonstrate that linear in $\gamma_{QPS}$ terms vanish identically in both expressions after averaging over the zero mode contained in the $\chi$-field. In order to evaluate the terms
$\sim \gamma_{QPS}^2$ it is convenient to make a shift $\chi_+(t)\to \Phi_0It+\chi_+(t)$ and to
decompose the averages by means of the Wick theorem. As a result we obtain
\begin{multline}
\langle V(t_1)\rangle = -\frac{\gamma_{QPS}^2}{e}\int dt\int\limits_0^L dx\int dt'\int\limits_0^L dx'\langle\dot \varphi_{+}(t_1) \chi_-(x,t) \rangle_0\\\times\langle\cos(\Phi_0I(t-t')+\chi_{+}(x,t)-\chi_{+}(x',t'))\cos(\chi_-(x,t)/2)\sin(\chi_-(x',t')/2) \rangle_0
\end{multline}
and
\begin{multline}
\langle V(t_1) V(t_2)\rangle = \frac{1}{4e^2}\langle\dot \varphi_{+}(t_1)\dot \varphi_{+}(t_2)\rangle_0
-\frac{\gamma_{QPS}^2}{2e^2}\int dt\int\limits_0^L dx\int dt'\int\limits_0^L dx'\langle\dot \varphi_{+}(t_1) \chi_{+}(x,t) \rangle_0\langle\dot\varphi_{+}(t_2)\chi_{-}(x,t) \rangle_0\\\times\langle
\sin(\Phi_0I(t'-t)+\chi_{+}(x',t')-\chi_{+}(x,t))\cos(\chi_-(x,t)/2)\sin(\chi_-(x',t')/2) \rangle_0
\\-\frac{\gamma_{QPS}^2}{2e^2}\int dt\int\limits_0^L dx\int dt'\int\limits_0^L dx'\langle\dot \varphi_{+}(t_1) \chi_{+}(x,t) \rangle_0\langle\dot\varphi_{+}(t_2)\chi_{-}(x',t') \rangle_0\\\times\langle
\sin(\Phi_0I(t'-t)+\chi_{+}(x',t')-\chi_{+}(x,t))\sin(\chi_-(x,t)/2)\cos(\chi_-(x',t')/2) \rangle_0
\\-\frac{\gamma_{QPS}^2}{2e^2}\int dt\int\limits_0^L dx\int dt'\int\limits_0^L dx'\langle\dot \varphi_{+}(t_1) \chi_{-}(x,t) \rangle_0\langle\dot\varphi_{+}(t_2)\chi_{+}(x,t) \rangle_0\\\times\langle
\sin(\Phi_0I(t'-t)+\chi_{+}(x',t')-\chi_{+}(x,t))\cos(\chi_-(x,t)/2)\sin(\chi_-(x',t')/2) \rangle_0
\\-\frac{\gamma_{QPS}^2}{2e^2}\int dt\int\limits_0^L dx\int dt'\int\limits_0^L dx'\langle\dot \varphi_{+}(t_1) \chi_{-}(x,t) \rangle_0\langle\dot\varphi_{+}(t_2)\chi_{+}(x',t') \rangle_0\\\times\langle
\sin(\Phi_0I(t-t')+\chi_{+}(x,t)-\chi_{+}(x',t'))\cos(\chi_-(x,t)/2)\sin(\chi_-(x',t')/2) \rangle_0
\\-\frac{\gamma_{QPS}^2}{4e^2}\int dt\int\limits_0^L dx\int dt'\int\limits_0^L dx'\langle\dot \varphi_{+}(t_1) \chi_{-}(x,t) \rangle_0\langle\dot\varphi_{+}(t_2)\chi_{-}(x',t') \rangle_0\\\times\langle
\cos(\Phi_0I(t-t')+\chi_{+}(x,t)-\chi_{+}(x',t'))\cos(\chi_-(x,t)/2)\cos(\chi_-(x',t')/2) \rangle_0
\end{multline}
The averages in Eqs. (S1) and (S2) are Gaussian and, hence, can be handled in a straightforward manner. After
that we immediately arrive at our final results for the $I-V$ curve (\ref{V1}), (\ref{P}) and for the
voltage noise spectrum (\ref{VV2})-(\ref{Sa}). Both these results are expressed via the function
$P_{x,x'}(\omega)$ (\ref{P}) which in turn contains the Green function $\mathcal G(x,x';t,0)$.

\subsection{Analytic structure of the Green functions}

Let us define a more general Green function $\mathcal G_{\chi}(x,x';\sigma)$ which depends on the complex time $\sigma$ and obeys the condition $\mathcal G_\chi(x,x';t-i0)=\mathcal G(x,x';t,0)$. With the aid of the Kubo-Martin-Schwinger condition one can deduce that the Green function $\mathcal G_{\chi}$ is periodic in the imaginary time direction, i.e.
\begin{equation}
\mathcal G_{\chi}(x,x';\sigma)=\mathcal G_{\chi}(x,x';\sigma-i/T).
\label{periodic}
\end{equation}
This function is analytic and has branch cuts at ${\rm Im}(\sigma)=N/T$ for all integer $N$. The function $\exp(i \mathcal G_\chi(x,x';\sigma))$ has the same analytic properties. One can write
\begin{equation}
 \mathcal P_{x,x'}(\omega)=\int\limits_{-\infty}^\infty dt e^{i\omega t}e^{i\mathcal G_{\chi}(x,x';t-i0)}.
\end{equation}
Distorting the integration path and utilizing the property $\mathcal G_\chi(x,x';\sigma)=-\bar{\mathcal G}_\chi(x,x';-\sigma)$ together with Eq. (\ref{periodic}) we arrive at the detailed balance condition (\ref{dbc}).

\subsection{Green functions }
The Green functions for the system displayed in Fig. 1 can be evaluated directly with the results:
\begin{equation}
G^R_{\varphi\varphi}(\omega)=\frac{1}{\frac{\omega^2}{2E_C}+\frac{i\omega}{4e^2R_x}-\frac{\omega\lambda}{\pi}\cot\left(\frac{\omega L}{v}\right)},
\end{equation}
\begin{equation}
G^R_{\chi\varphi}(x;\omega)=-G^R_{\varphi\chi}(x;\omega)=\frac{2i\lambda\cos\left(\frac{\omega (L-x)}{v}\right)}{\left(\frac{\omega^2}{2E_C}+\frac{i\omega}{4e^2R_x}\right)\sin\left(\frac{\omega L}{v}\right)-\frac{\omega\lambda}{\pi}\cos\left(\frac{\omega L}{v}\right)}
\label{23}
\end{equation}
and
\begin{multline}
G^R_{\chi\chi}(x,x';\omega)=\frac{4\pi\lambda\left(\cos\left(\frac{\omega(L-x)}{v}\right)\cos\left(\frac{\omega x'}{v}\right)\theta(x-x')+\cos\left(\frac{\omega (L-x')}{v}\right)\cos\left(\frac{\omega x}{v}\right)\theta(x'-x)
\right)}{\omega\sin\left(\frac{\omega L}{v}\right)}
\\+\frac{4\lambda^2\cos\left(\frac{\omega (L-x)}{v}\right)\cos\left(\frac{\omega(L- x')}{v}\right)}{\sin\left(\frac{\omega L}{v}\right)\left(\left(\frac{\omega^2}{2E_C}+\frac{i\omega}{4e^2R_x}\right)\sin\left(\frac{\omega L}{v}\right)-\frac{\omega\lambda}{\pi}\cos\left(\frac{\omega L}{v}\right)\right)}
\label{24}
\end{multline}
The last two expressions take a much simpler form in the long wire limit, in which case all plasmon excitations moving towards the grounded end of the wire eventually disappear and never pop up again while excitations moving in the opposite direction produce voltage fluctuations measured by a detector.
In this limit Eqs. (\ref{23}) and (\ref{24}) reduce to
\begin{equation}
G^R_{\varphi\chi}(x;\omega)\simeq -\frac{2\lambda e^{i\frac{\omega x}{v}}}{(\omega+i0)\left(\frac{\omega}{2E_C}+\frac{i\lambda}{\pi}\right)}, \quad
G^R_{\chi\chi}(x,x';\omega)\simeq -\frac{2\pi i\lambda }{\omega+i0}e^{i\frac{\omega|x-x'|}{v}}.
\end{equation}
Here we also set $R_x\to\infty$ as requested in the current bias limit.

In order to evaluate the general expressions for the $I-V$ curve (\ref{V1}), (\ref{P}) and for the
voltage noise (\ref{VV2})-(\ref{Sa}) it is necessary to compute the integral 
\begin{equation}
 \Upsilon(\omega,\Omega)=\int\limits_{-L/2}^{L/2} dx\int\limits_{-L/2}^{L/2}dx' e^{i\frac{\Omega}{v}(x-x')}\mathcal P_{x,x'}(\omega)
\end{equation}
Separating the left movers and the right movers, making use of the explicit form of the Green function $G^R_{\chi\chi}(x,x';\omega)$ and introducing the high frequency cutoff $\omega_c\sim1/\tau_0$ in order to avoid unphysical divergencies we obtain
\begin{equation}
 \Upsilon(\omega,\Omega)\simeq L\int\limits_{-\infty}^{\infty} dx e^{i\frac{\Omega}{v} x} \mathcal P_{x,0}(\omega)=\frac{Lv}{2}\varpi\left(\frac{\omega}{2}+\frac{\Omega}{2}\right)\varpi\left(\frac{\omega}{2}-\frac{\Omega}{2}\right),
\end{equation}
where
\begin{equation}
 \varpi(z)=\int\limits_{-\infty}^{\infty}dte^{izt}\frac{\sinh^\lambda(\pi T\tau_0)}{\sinh^{\lambda/2}(\pi T(\tau_0-t+i0))\sinh^{\lambda/2}(\pi T(\tau_0+t-i0))}.
\label{varpi}
\end{equation}
Performing the integration in Eq. (\ref{varpi}) we find
\begin{equation}
\varpi(\omega)=\frac{2^{\lambda}(\pi T\tau_0)^\lambda}{2\pi T}\frac{\Gamma\left(\frac{\lambda}{2}-\frac{i\omega}{2\pi T}\right)\Gamma\left(\frac{\lambda}{2}+\frac{i\omega}{2\pi T}\right)e^{\frac{\omega}{2T}}}{\Gamma(\lambda)}\equiv \varsigma(\omega)e^{\frac{\omega}{2T}}.
\end{equation}
The function $\varsigma(\omega)$ (\ref{vsi}) is directly employed in our results both for the $I-V$ curve (\ref{V3}) and for the QPS noise spectrum (\ref{shot1}), (\ref{Sa3}).

\end{widetext}

\begin{references}
\bibitem{FN0} For the sake of simplicity here we assume that $v\tau_0 \sim x_0$, where $x_0 \sim \xi$
is the QPS core size in space.
\bibitem{FN} Due to momentum conservation plasmons can only be created in pairs with the total zero momentum.
\end{references}
\end{document}